\documentclass[twocolumn]{jpsj3} %% two-column layout
\title{Quantitative Aspects of the Dynamical CPA in Harmonic Approximation}

\author{Toshihito \textsc{Tamashiro},
Shota \textsc{Nohara}, Keisuke \textsc{Miyagi}, 
and Yoshiro \textsc{Kakehashi}\thanks{E-mail address:
yok@sci.u-ryukyu.ac.jp, to be published in J. Phys. Soc. Jpn.}
}

\inst{
Department of Physics and Earth Sciences,
Faculty of Science, University of the Ryukyus, \\
1 Senbaru, Nishihara, Okinawa, 903-0213, Japan
}

\abst{
Magnetic and electronic properties of the Hubbard model on the Bethe and
fcc lattices in infinite dimensions have been investigated numerically 
on the basis of
the dynamical coherent potential approximation (CPA) theory combined
with the harmonic approximation (HA) in order to clarify the
quantitative aspects of the theory.  It is shown that the dynamical
CPA+HA reproduces well the sublattice magnetization, 
the magnetizations, susceptibilities, and 
the N\'eel temperatures 
($T_{\rm N}$) as well as the Curie temperatures calculated
by the Quantum Monte-Carlo (QMC) method.  The critical Coulomb
interactions ($U_{\rm c}$) for the metal-insulator (MI) transition
are also shown to agree with the QMC results above $T_{\rm N}$.  
Below $T_{\rm N}$, 
$U_{\rm c}$ deviate from the QMC values by about 30\% at low temperature 
regime.  These results indicate that the dynamical CPA+HA is
applicable to the quantitative description of the magnetic properties in
high dimensional systems, but one needs to take into account
higher-order dynamical corrections in order to describe the MI transition
quantitatively at low temperatures.
}

\kword{ Dynamical CPA, Curie temperature, N\'eel temperature,
Effective Bohr magneton number, Metal-insulator transition, 
Excitation spectra, Hubbard model
}

\begin{document}
\maketitle

\section{Introduction} 
The magnetism of correlated electron systems with intermediate
Coulomb interaction strength has been one of the long standing problems
in the solid-state physics because these systems show the complex
properties known as the itinerant vs. localized 
behaviors~\cite{fulde95,kake04}.  
The ground-state magnetizations of Fe, 
Co, Ni, for example, show the noninteger values in unit of the Bohr
magneton number being characteristic of a band model, 
while their finite-temperature properties such as
the magnetization vs. temperature curves and the Curie-Weiss
susceptibilities are well 
explained by a Heisenberg-type localized model.  The Hartree-Fock
theory and simple perturbation theories could not explain their magnetic 
properties since these theories overestimate the magnetic energy at
finite temperatures.  

Because of the difficulty in describing the itinerant vs. localized
behavior in the magnetism of transition metals, interpolation theories
have been developed by many investigators. 
Cyrot~\cite{cyrot72} adopted the functional integral method~\cite{mora74} 
to the Hubbard model and
derived the $P$-$T$ phase diagram of metal-insulator (MI) phase transition
at finite temperatures.  Hubbard~\cite{hub79} and Hasegawa~\cite{hase79} 
independently developed
the single-site spin fluctuation theory (SSF) on the basis of the
functional integral method.
The theory transforms the electron-electron interaction into a
time-dependent random potential by introducing the time-dependent 
charge and exchange random fictitious field variables, and treats 
the potentials using the coherent potential approximation 
(CPA)~\cite{ehren76}.  
The theory interpolates between the weak and
strong Coulomb interaction limits, and explained qualitatively the
localized vs. itinerant behaviors of magnetism in transition metals.

The SSF is based on the static approximation (SA) which neglects the
time dependence of the fictitious fields.  Though the SA is exact in the
high temperature limit, it reduces to the Hartree-Fock theory at the
ground state.  Thus it does not take into account the electron
correlations as found at the zero temperature by Gutzwiller~\cite{gutz63}, 
Hubbard~\cite{hub63}, and Kanamori~\cite{kana63}.
Kakehashi and Fulde~\cite{kake85} proposed a variational theory 
at finite temperatures
which takes into account the ground-state electron correlations, and
showed that the correlations at finite temperatures can reduce the
Curie temperatures obtained by the SSF by a factor of two. 

Towards more quantitative theory, 
Kakehashi~\cite{kake92} proposed the dynamical CPA which completely 
takes into account
electron correlations at finite temperatures within the single-site
approximation, and clarified the basic properties of the theory using a
Monte-Carlo technique.  Later he proposed an analytic theory of the
dynamical CPA combined with the harmonic approximation
(HA)~\cite{kake02}, 
and examined the dynamical effects on the itinerant ferromagnetism.
In the HA~\cite{amit71}, 
we neglect the mode-mode couplings between the dynamical
potentials, and take into account the contributions from the dynamical
potentials with the same frequency independently.  The HA is known to
be exact up to the second order in the weak Coulomb interaction limit,
and is known to describe quantitatively the Kondo limit in the strong 
Coulomb interaction regime~\cite{dai91}.  
The dynamical CPA has recently been proved~\cite{kake04,kake02-2} 
to be equivalent to the many-body CPA in the disordered 
problem~\cite{hiro77}, 
the dynamical mean-field theory in the metal-insulator transition in
infinite dimensions~\cite{geor96}, 
and the projection-operator method CPA for excitations 
in solids~\cite{kake04-2}.

More recently, we proposed the first-principles dynamical 
CPA~\cite{kake08} 
on the basis of the tight-binding linear muffintin orbital (TB-LMTO) 
Hamiltonian~\cite{ander75,ander94} 
with the LDA (Local Density Approximation)+$U$
type intraatomic Coulomb interactions~\cite{anis97}, 
and clarified the ferromagnetism
of Fe, Co, and Ni~\cite{kake09,kake10}, 
as well as the systematic change of excitation spectra in 3$d$ 
transition metal series~\cite{kake10-2,kake11}.

Although the numerical results of the first-principles dynamical 
CPA combined with the HA seem to be reasonable and explained well 
a systematic change of the XPS and BIS data~\cite{kake10-2}, 
the validity of the 
dynamical CPA+HA has not yet been examined in details.  
The purpose of the present paper is to clarify the quantitative 
aspects of the dynamical CPA+HA from the numerical point of view.  
For this purpose, we have calculated the
antiferromagnetic properties of the half-filled
Hubbard model on the Bethe lattice in infinite dimensions~\cite{geor96} 
and the ferromagnetic properties of the non half-filled Hubbard model 
on the fcc lattice in infinite dimensions~\cite{mull91} 
on the basis of the dynamical CPA+ HA.  
We clarify the validity of the theory by
comparing the results of calculations with those of the Quantum 
Monte-Carlo (QMC) method~\cite{ulmke95,held97,schl99,moe95,ulmke98}.  
We will conclude
that the dynamical CPA approach is useful for the quantitative 
understanding of the magnetic properties of solids in high dimensions
where the single-site approximation works well.

Outline of the present paper is as follows.  In \S 2, we summarize
the dynamical CPA combined with the HA and extend the theory to the
antiferromagnetic case with the bipartite lattice.  In \S 3.1 we
present the results of calculations for the antiferromagnetic states
in the half-filled Hubbard model on the Bethe lattice.
It is shown that the zeroth approximation to the dynamical CPA, {\it
i.e.}, the SA quantitatively describes the N\'eel
temperatures ($T_{\rm N}$) as well as the MI boundary above $T_{\rm N}$.
We present in \S 3.2 the results of calculations for the densities of
states (DOS) as well as the MI transition below $T_{\rm N}$.
We will show that the dynamical effects tend to increase the critical
Coulomb interaction for the MI transition.
The numerical results for the ferromagnetic properties
on the fcc lattice in infinite dimensions are presented in \S 3.3.
The Curie temperatures as well as the paramagnetic susceptibilities are 
shown to be described quantitatively by means of
the dynamical CPA+HA in the quarter-filled regime of electron number.  
In the last \S 4 we summarize our numerical results.  
It is emphasized that the dynamical CPA is applicable to the 
quantitative description of the magnetic properties in high dimensions.

\section{Dynamical CPA in the ferro- and antiferro-magnetic states}
We adopt in the present paper the Hubbard model as follows.
\begin{eqnarray}
\hat{H} = \sum_{i,\sigma} \epsilon_{0} n_{i\sigma}
+ \sum_{i, j, \sigma} t_{i j} a_{i \sigma}^{\dagger} a_{j \sigma}
+ \sum_{i} U n_{i \uparrow}n_{i \downarrow} \ .
\label{hub-h}
\end{eqnarray}
Here $\epsilon_{0}$ is the atomic energy level, $t_{ij}$ is the transfer
integrals between sites $i$ and $j$, and $U$ is the intraatomic
Coulomb interaction energy parameter. 
$a_{i \sigma}^{\dagger} (a_{i \sigma})$ denotes the creation 
(annihilation) operator for an electron with spin $\sigma$ on site $i$,
and $n_{i\sigma}=a_{i \sigma}^{\dagger} a_{j \sigma}$ is the number
operator for the electron.

In the dynamical CPA~\cite{kake02}, 
we transform the two-body interaction 
$U n_{i \uparrow}n_{i \downarrow}$ in the free energy into a dynamical
one-body potential $v_{i}$ with time-dependent random spin and
charge fields using the Hubbard-Stratonovich 
transformation~\cite{hub59}.  
Then introducing the site-diagonal effective potential $\Sigma_{i}$
({\it i.e.}, the coherent potential) into the potential part of 
the free energy, we expand the correction
terms containing $v_{i} - \Sigma_{i}$ with respect to the site.
The zeroth term is the coherent term $\tilde{\mathcal{F}}[\Sigma]$ 
which does not depend on the dynamical potential at all. 
The next term consists of the single-site terms each of which
contains the dynamical potential on the same site.  
The higher-order terms $\Delta \mathcal{F}$ describe the inter-site 
spin and charge fluctuations.

In the dynamical CPA, we adopt the single-site approximation which
neglects the higher-order inter-site corrections $\Delta \mathcal{F}$.  
The free energy of the dynamical CPA is then written as follows.
\begin{eqnarray}
{\mathcal F}_{\rm CPA} = \tilde{\mathcal{F}}[\Sigma] 
- \beta^{-1} \sum_{i} {\rm ln} \int \sqrt{\frac{\beta U}{4\pi}}d\xi_{i} 
\ e^{\displaystyle -\beta E^{(i)}(\xi_{i})}
\  .
\label{fcpa3}
\end{eqnarray}
Here $\beta$ denotes the inverse temperature, $\xi_{i}$ is the static
exchange field on site $i$, and $E^{(i)}(\xi_{i})$ is a single-site 
effective potential projected onto the static field $\xi_{i}$.

The effective potential $E^{(i)}(\xi)$ consists of the
static part $E^{(i)}_{\rm st}(\xi)$ and the dynamical part
$E^{(i)}_{\rm dyn}(\xi)$.
\begin{eqnarray}
E^{(i)}(\xi) = E^{(i)}_{\rm st}(\xi) + E^{(i)}_{\rm dyn}(\xi) \ .  
\label{eimp3}
\end{eqnarray}
The static potential $E^{(i)}_{\rm st}(\xi)$ is obtained by
neglecting the time-dependence of the field variables and is given by
\begin{eqnarray}
E^{(i)}_{\rm st}(\xi) \!\!\!\!\!\! & = & \!\!\!\!\!\! 
-\frac{1}{\beta} \sum_{l,\sigma} \ln \big[
1 - (v^{(0)}_{i\sigma}(\xi) - \Sigma_{i\sigma}(i\omega_{l}) )
F_{i\sigma}(i\omega_{l}) \big]   \nonumber \\
& & - \frac{1}{4} U (\tilde{n}_{i}(\xi)^{2} - \xi^{2}) \ ,
\label{est}
\end{eqnarray}
\begin{eqnarray}
v^{(0)}_{i\sigma}(\xi) = \epsilon_{0} - \mu 
+ \frac{1}{2} U \tilde{n}_{i}(\xi) - \frac{1}{2} U \xi\sigma . 
\label{stlevel}
\end{eqnarray}
Here $v^{(0)}_{i\sigma}(\xi)$ is the Hartree-Fock type static potential.
$\mu$ denotes the chemical potential.  $\Sigma_{i\sigma}(i\omega_{l})$
is the frequency representation of a time-dependent coherent potential 
$\Sigma_{i\sigma}(\tau)$, $\tau$ being the imaginary time.
$\omega_{l}=(2l+1)\pi/\beta$ denotes the Matsubara frequency.
The electron number $\tilde{n}_{i}(\xi)$ for a given exchange field 
$\xi$ is defined by
\begin{eqnarray}
\tilde{n}_{i}(\xi) = \frac{1}{\beta} \sum_{l,\sigma} 
G_{i\sigma}(i\omega_{l}, \xi) \ .
\label{saddlept}
\end{eqnarray}
The Green function $G_{i\sigma}(i\omega_{l}, \xi)$ will 
be determined later self-consistently (see eq. (\ref{effimpg})).

The coherent Green function $F_{i\sigma}(i\omega_{l})$ in
eq. (\ref{est}) is defined by
\begin{eqnarray}
F_{i\sigma}(i\omega_{l}) = \big[
(i\omega_{l} - \boldsymbol{H}_{0} 
- \boldsymbol{\Sigma}_{\sigma}(i\omega_{l}))^{-1} \big]_{ii} \ .
\label{cohg}
\end{eqnarray}
Here $(\boldsymbol{H}_{0})_{ij} = (\epsilon_{0} - \mu)\delta_{ij} 
+ t_{ij}(1-\delta_{ij})$ is the one-electron Hamiltonian matrix element 
in eq. (\ref{hub-h}) and 
$(\boldsymbol{\Sigma}_{\sigma}(i\omega_{l}))_{ij} = 
\Sigma_{i\sigma}(i\omega_{l})\delta_{ij}$.

The dynamical potential $E^{(i)}_{\rm dyn}(\xi)$ in eq. (\ref{eimp3}) 
is given in the harmonic approximation (HA) as
\begin{eqnarray}
E^{(i)}_{\rm dyn}(\xi) = -\frac{1}{\beta} \ {\rm ln} \ 
\left[ 1 + \sum_{\nu=1}^{\infty} \sum_{n=1}^{\infty} 
\overline{D}^{(n)}_{i\nu} \right] \ ,
\label{edyn3}
\end{eqnarray}
\begin{eqnarray}
\overline{D}^{(n)}_{i\nu} = 
U^{2n} \left( \frac{\mathstrut i}{\mathstrut 2\pi\nu} \right)^{2n}
B^{(n)}_{i\nu\uparrow}B^{(n)}_{i\nu\downarrow} \ .
\label{dnudnu}
\end{eqnarray}
First few terms of $B^{(n)}_{i\nu\sigma}$ are given as 
follows~\cite{kake02}.
\begin{eqnarray}
B^{(1)}_{i\nu\sigma} = \frac{2\pi i \nu}{\beta} \sum_{l=0}^{\infty}
\Bigl( \tilde{g}_{i\sigma}(l-\nu)\tilde{g}_{i\sigma}(l) 
\hspace{20mm} \nonumber \\
+ \tilde{g}_{i\sigma}(l+\nu)^{\ast}\tilde{g}_{i\sigma}(l)^{\ast}  
\Bigr) \ , 
\label{extblnu1}
\end{eqnarray}
\begin{eqnarray}
B^{(2)}_{i\nu\sigma}  =  \left( \frac{2\pi\nu}{\beta} \right)^{2} {\rm Re}
\Biggl[ 2 \sum_{l=0}^{\infty} \tilde{g}_{i\sigma}(l-\nu)\tilde{g}_{i\sigma}(l)
\hspace{10mm} \nonumber \\
 \times \,\bigl( \tilde{g}_{i\sigma}(l-2\nu)\tilde{g}_{i\sigma}(l-\nu)+
\tilde{g}_{i\sigma}(l-\nu)\tilde{g}_{i\sigma}(l) \ \ \ \  \nonumber  \\
+ \tilde{g}_{i\sigma}(l)\tilde{g}_{i\sigma}(l+\nu) \bigr) \nonumber \\
 - \sum_{l=0}^{\nu -1} 
(\tilde{g}_{i\sigma}(l-\nu)\tilde{g}_{i\sigma}(l))^{2} 
\Biggr] + {B^{(1)}_{i\nu\sigma}}^{2} \ . \hspace{0mm}
\label{extblnu2}
\end{eqnarray}
Here $\tilde{g}_{i\sigma}(l)$ is the static Green function on site $i$, 
which is defined by
\begin{eqnarray}
\tilde{g}_{i\sigma}(l) = \big[ F_{i\sigma}(i\omega_{l})^{-1} - 
v^{(0)}_{i\sigma}(\xi) 
+ \Sigma_{i\sigma}(i\omega_{l}) \big]^{-1} \ .
\label{gst}
\end{eqnarray}

For the higher-order terms of $B^{(n)}_{i\nu\sigma}$, one can adopt the
following form of the asymptotic approximation.
\begin{eqnarray}
B^{(n)}_{i\nu\sigma} = \sum_{\textstyle \sum_{k=0}^{\nu-1}l_{k}=n} 
\frac{\displaystyle n\, !}
{\displaystyle  \Bigl[\prod_{k=0}^{\nu-1}l_{k}!\Bigr]}
\left[\prod_{k=0}^{\nu-1} B_{i\nu\sigma}^{(l_{k})}(k) \right] \ ,
\label{blnu}
\end{eqnarray}
\begin{eqnarray}
B^{(l)}_{i\nu\sigma}(k) & = & 
b^{(0)}_{l\sigma}(\nu,k) 
+ \sum_{m=0}^{l-1} (-)^{l-m} \biggl( \!\!
\begin{array}{l}
\,l \vspace{-2mm} \\
m  \end{array}
\!\! \biggr)  \nonumber \\
& & \hspace*{-8mm} \times \Big[ \ 
b^{(0)}_{m\sigma}(\nu,k) b^{(0)}_{l-m \sigma}(-\nu,k) 
 \nonumber \\
& + &
\frac{2\pi(l-m)\nu}{i\beta}\tilde{g}_{i\sigma}(-\nu+k)\tilde{g}_{i\sigma}(k)
 \nonumber \\
& & \ \ \ \times b^{(1)}_{m\sigma}(\nu,k) b^{(1)}_{l-m-1 \sigma}(-\nu,k)
\Big] \ . 
\label{blnuasym}
\end{eqnarray}
The functions $b^{(0)}_{m\sigma}(\pm\nu,k)$ and 
$b^{(1)}_{m\sigma}(\pm\nu,k)$ in the $0$-th order and the second order
asymptotic approximation are given in Appendix B of Ref.~\cite{kake02}.

The coherent potential $\Sigma_{i\sigma}(i\omega_{l})$ is determined so
that the higher-order inter-site corrections become minimum.  The
condition called the CPA equation is given by 
\begin{eqnarray}
\left\langle G_{i\sigma}(i\omega_{l}, \xi) \right\rangle_{\rm eff}
= F_{i\sigma}(i\omega_{l}) \ .
\label{cpa2}
\end{eqnarray}
Here the average $\langle \ \rangle_{\rm eff}$ 
at the l.h.s. (left-hand-side) of the above equation means taking 
a classical average with respect to the effective potential
$E^{(i)}(\xi)$.  The on-site dynamical impurity Green function is given
by 
\begin{eqnarray}
G_{i\sigma}(i\omega_{l},\xi) = 
\tilde{g}_{i\sigma}(l)
+ \frac{\displaystyle 
\sum_{\nu=1}^{\infty}\sum_{n=1}^{\infty} 
\frac{\displaystyle \delta \overline{D}^{(n)}_{i\nu}}
{\displaystyle \kappa_{i\sigma}(i\omega_{l})
\delta\Sigma_{i\sigma}(i\omega_{l})}
}
{\displaystyle 1 + 
\sum_{\nu=1}^{\infty} \sum_{n=1}^{\infty} \overline{D}^{(n)}_{i\nu}
} \ .
\label{effimpg}
\end{eqnarray}
Here $\kappa_{i\sigma}(i\omega_{l}) = 1- F_{i\sigma}(i\omega_{l})^{-2}
\delta F_{i\sigma}(i\omega_{l}) / \delta\Sigma_{i\sigma}(i\omega_{l})$.

The average electron number $\langle n_{i} \rangle$ and the local
magnetic moment $\langle m_{i} \rangle$ are given by
\begin{eqnarray}
\langle n_{i} \rangle = \langle \tilde{n}_{i}(\xi_{i}) 
\rangle_{\rm eff} \ ,
\label{n3}
\end{eqnarray}
\begin{eqnarray}
\langle m_{i} \rangle = \langle \xi_{i} \rangle_{\rm eff} \ .
\label{m3}
\end{eqnarray}
The double occupation number $\langle n_{i\uparrow}n_{i\downarrow} \rangle$
and the amplitudes of local moment on site $i$ are obtained from 
the following expressions~\cite{kake02}.  
\begin{eqnarray}
\langle n_{i\uparrow}n_{i\downarrow} \rangle =
\frac{1}{4}\langle \tilde{n}_{i}(\xi_{i})^{2} \rangle_{\rm eff} 
- \frac{1}{4} \left( \langle \xi^{2}_{i} \rangle_{\rm eff} 
- \frac{2}{\beta U} \right)  \nonumber \\
+ \left\langle \left[ 
\frac{\mathstrut \partial E^{(i)}_{\rm dyn}(\xi_{i})}
{\mathstrut \partial U} \right]_{v} 
\right\rangle_{\rm eff} \ ,
\label{ndble}
\end{eqnarray}
\begin{eqnarray}
\langle m^{2}_{i} \rangle = \langle n_{i} \rangle 
- 2 \langle n_{i\uparrow}n_{i\downarrow} \rangle \ .
\label{m23}
\end{eqnarray}
Here $[\ ]_{v}$ means to take the derivative of a quantity fixing the
static potential $v^{(0)}_{i\sigma}(\xi_{i})$.

Equations (\ref{cpa2}) and (\ref{n3}) form a set of self-consistent
equations to determine the coherent potential 
\{$\Sigma_{i\sigma}(i\omega_{l})$\} and the chemical potential  
$\epsilon_{0}-\mu$.
However it is time consuming to solve the CPA equation (\ref{cpa2}) 
because one has to take the average $\langle \ \rangle_{\rm eff}$ 
at each frequency $\omega_{l}$.
The following decoupling approximation simplifies the numerical
calculations.
\begin{eqnarray}
\sum_{q=\pm 1} \frac{1}{2} 
\left( 1 + q \frac{\langle \xi_{i} \rangle_{\rm eff}}{x_{i}} \right)
G_{i\sigma}(i\omega_{l}, qx_{i}) 
= F_{i\sigma}(i\omega_{l}) \ .
\label{cpa3}
\end{eqnarray}
Here $x_{i} = \sqrt{\langle \xi^{2}_{i} \rangle_{\rm eff}}$. 
Note that the approximation is correct up to 
the second moment ({\it i.e.}, $\langle \xi_{i} \rangle_{\rm eff}$ and 
$\langle \xi^{2}_{i} \rangle_{\rm eff}$).  

We assume in this work that each site is crystallographically
equivalent to each other.  In the para- and ferro-magnetic
states, we can assume a site-independent coherent potential 
$\Sigma_{\sigma}(i\omega_{l})$.
The site-independent coherent Green function $F_{\sigma}(i\omega_{l})$ 
is then given by 
\begin{eqnarray}
F_{\sigma}(i\omega_{l}) = \int \frac{\mathstrut \rho(\epsilon)d\epsilon}
{i\omega_{l}-\epsilon_{0}+\mu-\Sigma_{\sigma}(i\omega_{l})-\epsilon} \ .
\label{cohgp}
\end{eqnarray}
Here $\rho(\epsilon)$ is the density of states (DOS) for the
noninteracting Hamiltonian matrix $t_{ij}$.

In the antiferromagnetic state with sublattice magnetization, we have
two types of coherent potential; $\Sigma^{(+)}_{\sigma}(i\omega_{l})$ on
the up-spin sublattice and $\Sigma^{(-)}_{\sigma}(i\omega_{l})$ on the
down-spin sublattice.  Because of the symmetric relation 
$\Sigma^{(-)}_{\sigma}(i\omega_{l})=\Sigma^{(+)}_{-\sigma}(i\omega_{l})$, 
it is enough to require the self-consistency of 
$\Sigma^{(+)}_{\sigma}(i\omega_{l})$ on the up-spin sublattice.  
The corresponding coherent Green function
$F^{(+)}_{\sigma}(i\omega_{l})$ is given by~\cite{foo70,plis73} 
\begin{eqnarray}
F^{(+)}_{\sigma}(i\omega_{l}) = 
\sqrt{\dfrac{i\omega_{l}\!-\!\Sigma^{(+)}_{-\sigma}(i\omega_{l})}
{i\omega_{l}\!-\!\Sigma^{(+)}_{\sigma}(i\omega_{l})}} \hspace{33mm}
\nonumber \\
\times \int \!\! \frac{\mathstrut \rho(\epsilon)d\epsilon}
{\sqrt{(i\omega_{l}\!-\!\Sigma^{(+)}_{+}(i\omega_{l}))
(i\omega_{l}\!-\!\Sigma^{(+)}_{-}(i\omega_{l}))} - 
\! \epsilon_{0} \! + \! \mu \! - \! \epsilon} \ .
\label{cohgaf}
\end{eqnarray}

We adopted eqs. (\ref{cohgp}) and (\ref{cohgaf}) in the para- and
antiferro-magnetic calculations on the Bethe lattice and in the
ferromagnetic calculations on the fcc lattice in infinite dimensions.
Moreover in the self-consistent calculations with use of the HA, we 
adopted eqs. (\ref{extblnu1}) and (\ref{extblnu2}) for $U^{2}$ and 
$U^{4}$ terms and took into account the higher-order terms up to
$U^{16}$ using the asymptotic approximation, {\it i.e.} 
eqs. (\ref{blnu}) and (\ref{blnuasym}).    
%
%
%---------------------------------------------------------------------
\begin{figure}
\includegraphics[scale=0.7]{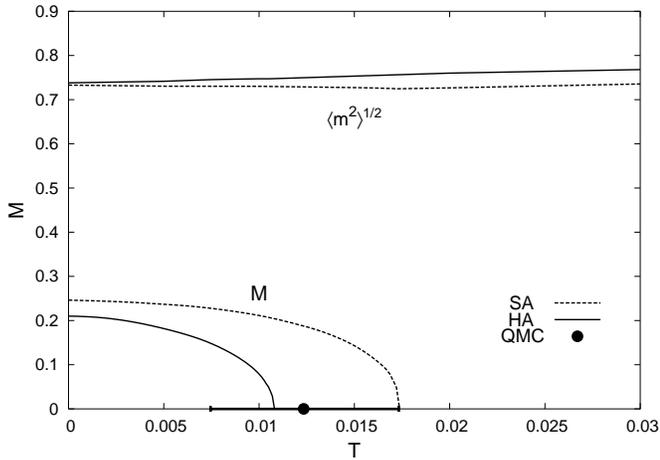}%
\caption{\label{figmtu0.5}
Sublattice magnetization vs. temperature curves and the amplitude of 
local moments $\langle m^{2} \rangle^{1/2}$ for the half-filled
 Hubbard model on the Bethe lattice in infinite dimensions.
Coulomb interaction energy parameter is chosen to be $U=0.5$ 
in unit of half the band width for noninteracting density of states. 
Dashed curves show the results of the static approximation (SA), 
solid curves show the results of the harmonic approximation (HA), 
closed circle with error bar indicates the N\'eel temperature 
obtained by the Quantum 
Monte-Carlo (QMC) method~\cite{ulmke95,held97}.
}
\end{figure}
%---------------------------------------------------------------------

\section{Numerical Results}

\subsection{Antiferromagnetic states on the Bethe lattice}
We present in this subsection the numerical results of calculations 
for the antiferromagnetic states at half filling on the Bethe lattice in
infinite dimensions.  The DOS for the noninteracting system is given by
a semi-elliptical form as
\begin{eqnarray}
\rho(\epsilon) = \dfrac{2}{\pi W^{2}} \sqrt{W^{2} - \epsilon^{2}} \ .
\label{bdos}
\end{eqnarray}
Here $2W$ denotes the band width.
In the present and next subsections, we adopt the energy unit to 
be $W=1$.
%---------------------------------------------------------------------
\begin{figure}
\includegraphics[scale=0.7]{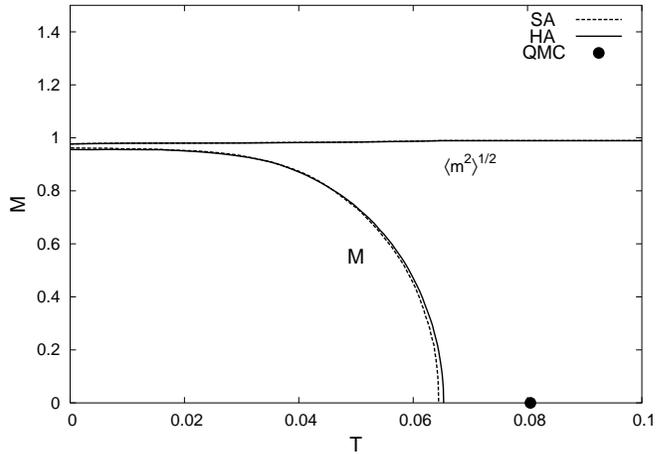}%
\caption{\label{figmtu3.5}
Sublattice magnetization vs. temperature curves and the amplitude of 
local moments for $U=3.5$. 
Notations are the same as in Fig. 1.
}
\end{figure}
%---------------------------------------------------------------------

We found that the antiferromagnetism is stabilized at half filling with
decreasing temperatures.  The numerical results of sublattice
magnetization vs. temperature curves are presented in Fig. 1 
for rather small Coulomb interaction energy parameter $U=0.5$.  
Calculated N\'eel
temperatures $T_{\rm N}$ are 0.017 in the static approximation (SA)
and 0.011 in the harmonic approximation (HA), respectively.
The dynamical effects reduce the ground-state sublattice magnetization
(extrapolated value) by 15\% and $T_{\rm N}$ by 36\%.  The calculated
N\'eel temperature with dynamical corrections seems to be in good 
agreement with the QMC result as shown in Fig. 1.
Although the agreement might not be so convincing because of a large
error bar in the QMC calculations, quantitative description in the weak
Coulomb interaction regime is expected because the HA is exact up 
to the second order in $U$ in the weak Coulomb interaction limit.  
The amplitude of local moment $\langle m^{2} \rangle^{1/2}$
is also enhanced by the dynamical effects because electron
correlations suppress the double occupancy of electrons even 
at finite temperatures.

The results for a strong Coulomb interaction $U=3.5$ are presented in
Fig. 2.  The sublattice magnetization and N\'eel temperature are much
enhanced as compared with the case of $U=0.5$.  The ground-state
sublattice magnetization (0.960 $\mu_{\rm B}$ for the SA
and 0.955 $\mu_{\rm B}$ in the HA) is close to the atomic value 1.0 
$\mu_{\rm B}$.
Calculated $T_{\rm N}$ are 0.0645 in the SA and 0.0654 in the HA, 
respectively; the $T_{\rm N}$
is slightly increased by the dynamical effects, but is smaller than the
QMC value~\cite{ulmke95} by 18\%.
%---------------------------------------------------------------------
\begin{figure}
\includegraphics[scale=0.7]{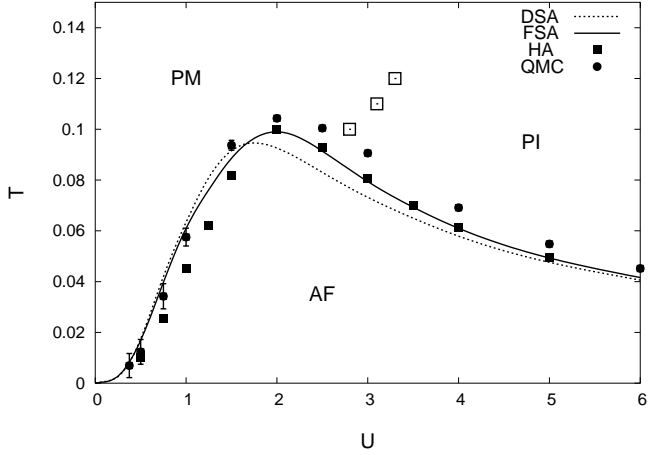}%
\caption{\label{figneel}
N\'eel temperatures ($T_{\rm N}$) in the SA with decoupling 
approximation (dotted curve), the full SA 
(solid curve), the HA with dynamical 
corrections (closed squares), and the QMC method 
(closed circles)~\cite{ulmke95,held97}. The open squares around $U=3.0$ 
above $T_{\rm N}$ indicate the metal-insulator crossover points 
in the full SA.
Below $T_{\rm N}$, the antiferromagnetic state (AF) is stabilized, while
 above $T_{\rm N}$ the paramagnetic metal (PM) and the paramagnetic
 insulator (PI) are realized.
}
\end{figure}
%---------------------------------------------------------------------

We have calculated the N\'eel temperatures for various Coulomb
interaction energy parameters.  The results are summarized in Fig. 3.
It should be noted that we adopted the decoupling approximation 
(\ref{cpa3}) in the calculations of Figs. 1 and 2 in order to
reduce the computation time.  To clarify the difference between the
decoupling approximation (\ref{cpa3}) and the full self-consistent CPA 
(\ref{cpa2}), we performed the full static calculations with use of 
(\ref{cpa2}).  The calculated $T_{\rm N}$ vs. $U$ curve is shown by solid
curve and is compared with those in the decoupling approximation as well
as the QMC results.  In both the weak and strong Coulomb interaction
regimes ({\it i.e.}, $U < 1$ and $U > 3$), 
calculated $T_{\rm N}$ with different approximation schemes 
yield basically the same result.
In the intermediate Coulomb interaction regime ({\it i.e.}, $1 < U < 3$), 
the N\'eel temperatures in the decoupling
approximation deviate from the full static results by several percent.
It should be noted that $T_{\rm N}$ calculated by the full SA 
quantitatively agree with the QMC results.  This indicates
that the SA provides us with a good starting point to
calculate quantitatively the magnetic properties of electrons 
especially at high temperatures. 

We have estimated the N\'eel temperatures in the full HA by adding 
$\Delta T_{\rm N} = T_{\rm N}({\rm HA}) - T_{\rm N}({\rm SA})$
in the decoupling approximation to $T_{\rm N}$ in the full static 
one~\cite{kcmmt}.
The results are shown in Fig. 3 by closed squares.  We find
that the dynamical effects in the harmonic approximation reduce 
$T_{\rm N}$ in the weak Coulomb interaction regime ($U < 1$) 
and slightly increase 
$T_{\rm N}$ in the strong Coulomb interaction regime ($U > 2$).  
Calculated $T_{\rm N}$ are in agreement with the QMC results in the
weak $U$ regime, but are underestimated in the intermediate and strong 
$U$ regimes ($1 \lesssim U$).  In particular the present dynamical 
calculations hardly correct $T_{\rm N}$ in the SA 
in the strong $U$ regime ($2 \lesssim U$).
One of the possible reasons for insufficient corrections is
that the decoupling approximation is not suitable at high temperatures; 
it tends to overestimate the magnetic entropy at high temperatures 
because the approximation implies to replace a broad distribution 
$p(\xi)=\exp (-\beta E^{(i)}(\xi))/\int d\xi \exp (-\beta E^{(i)}(\xi))$ 
with a two-delta function.  
The higher order dynamical corrections should also be taken into account 
in eq. (\ref{dnudnu}) for strong $U$ regime in order to make more 
reasonable corrections.
%---------------------------------------------------------------------
\begin{figure}
\includegraphics[scale=0.7]{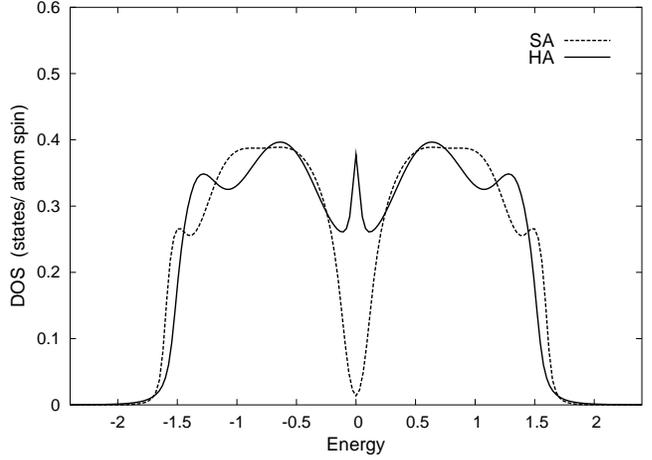}%
\caption{\label{figdosm}
Density of states (DOS) for $U=1.7$ and $T=0.06$ in the SA 
(dashed curve), and in the HA with dynamical corrections (solid curve).
}
\end{figure}
%---------------------------------------------------------------------

\subsection{Metal-insulator transition}
Single-particle excitation spectra are obtained by means of 
a numerical analytic continuation~\cite{vid77} 
of the self-energy for temperature 
Green function, {\it i.e.}, the coherent potential 
$\Sigma_{i\sigma}(i\omega_{l})$.  It should be noted that
the coherent potential obtained by solving the CPA equation (\ref{cpa3})
in the decoupling approximation is an approximate solution to the full 
equation (\ref{cpa2}).  In order to obtain accurately the DOS as the
single-particle excitations, we improved the coherent potential adopting
the average $t$-matrix approximation (ATA)~\cite{ehren76,korr58} 
after we solved eq. (\ref{cpa3}).
%---------------------------------------------------------------------
\begin{figure}
\includegraphics[scale=0.7]{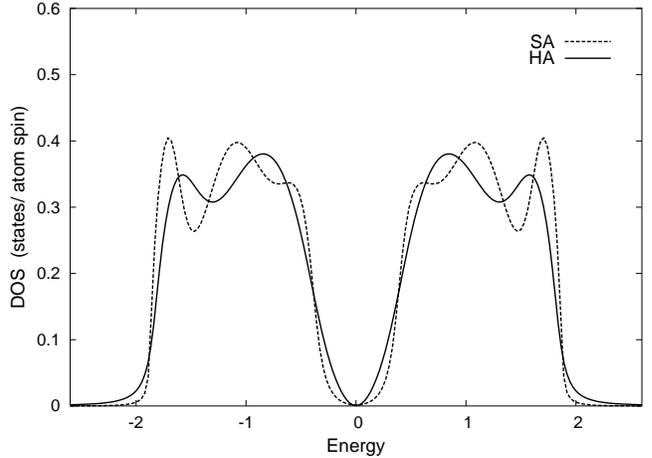}%
\caption{\label{figdosi}
DOS for $U=2.2$ and $T=0.06$ in the SA (dashed curve), and in the HA 
with dynamical corrections (solid curve).
}
\end{figure}
%---------------------------------------------------------------------
%
%
\begin{eqnarray}
\Sigma^{\rm ATA}_{i\sigma}(i\omega_{n}) \! = \! 
\Sigma_{i\sigma}(i\omega_{n}) \! + \! 
\dfrac{\langle G_{i\sigma}(i\omega_{n}, \xi)
\rangle_{\rm eff} \! - \! F_{i\sigma}(i\omega_{n})}
{\langle G_{i\sigma}(i\omega_{n}, \xi) \rangle_{\rm eff}
F_{i\sigma}(i\omega_{n})} .
\label{ata}
\end{eqnarray}
Here the coherent potential in the decoupling approximation was used at
the r.h.s. (right-hand-side) of the above equation, but the full average 
$\langle \ \rangle_{\rm eff}$ of the impurity Green function is taken.  
The ATA is a one-shot correction to the full CPA (\ref{cpa2}).

Making use of the Pad\'e numerical analytic continuation~\cite{vid77}, 
we obtained $\Sigma^{\rm ATA}_{i\sigma}(\omega+i\delta)$.  
Here $\omega$ is an energy variable
on the real axis and $\delta$ is an infinitesimal
positive number.  Using the self-energy, we obtained the coherent Green
function $F^{(+)}_{\sigma}(\omega+i\delta)$, and calculated the DOS via the
relation,
\begin{eqnarray}
\rho_{\sigma}(\omega) = - \frac{1}{\pi} {\rm Im} 
\, F^{(+)}_{\sigma}(\omega+i\delta) \ .
\label{fdos}
\end{eqnarray}

The system changes from metal to insulator with the formation of a gap
on the Fermi level with increasing Coulomb interaction.  In the full
SA, the gap is gradually formed above the N\'eel
temperatures.  The crossover points are shown in Fig. 3 by open squares.
When we assume the paramagnetic state below $T_{\rm N}$, the formation
of a gap becomes clearer at lower temperatures, indicating the
metal-insulator (MI) transition.

Figure 4 shows the DOS in the metallic state near the MI transition.
We find the upper and lower Hubbard bands.  A gap is
almost opened on the Fermi level in the SA, while the
quasiparticle peak remain in the dynamical calculations.
When the Coulomb interaction energy $U$ is increased, a gap is opened as
shown in Fig. 5.  The dynamical effects shift the spectral weight to the
lower energy region.
%---------------------------------------------------------------------
\begin{figure}
\includegraphics[scale=0.7]{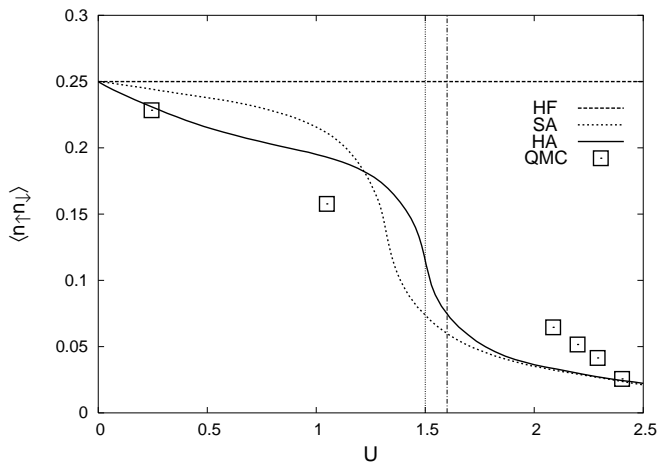}%
\caption{\label{figdble}
Double occupation numbers as a function of $U$ at $T=0.02$ in the 
Hartree-Fock (HF) approximation (dashed curve),  the SA (dotted curve), 
and the HA (solid curve). The paramagnetic state is 
assumed. Open squares show the double occupation numbers in the QMC 
calculations~\cite{geor96} at $T=0.03125$. The vertical lines indicate 
the critical Coulomb interaction $U_{\mathrm{c}}$ in the SA 
($U_{\rm c}=1.5$: thin dotted line) and HA ($U_{\rm c}=1.6$: 
dot-dashed line).
Both values of $U_{\mathrm{c}}$ are obtained within the decoupling 
approximation (\ref{cpa3}). 
}
\end{figure}
%---------------------------------------------------------------------

The metal-insulator transition is accompanied by a localization of
electrons.  We show in Fig. 6 an example of double occupation number
$\langle n_{\uparrow}n_{\downarrow} \rangle$ vs $U$ curve at low temperatures.
The Hartree-Fock approximation yields the constant value 0.25
irrespective of $U$ because of no on-site correlations.  
The double occupation
number in the SA reduces to the Hartree-Fock value in
the weak Coulomb interaction limit.  It rapidly decreases near the MI
transition, and gives the alloy-analogy results of the Hubbard 
theory~\cite{hub63} 
in the strong $U$ limit.  The dynamical effects based on the HA reduce 
further the double occupancy in the weak $U$ regime and
enhance it in the strong $U$ regime as shown in Fig. 6.
  
There are no QMC calculations on the double occupation number at
$T=0.02$.  We plotted in Fig. 6 the QMC results~\cite{geor96} at 
$T=0.031$ for qualitative or semiquantitative comparison with the 
present results.
The QMC value at $U=0.25$ indicates that the dynamical CPA+HA yields the
quantitative result in the weak Coulomb interaction.
The QMC result at $U=1.0$ suggests that the HA is not sufficient to
reduce the double occupancy in the intermediate regime of the Coulomb
interaction.  It should be noted that the critical Coulomb interaction
in the QMC is $U_{\rm c}=2.4$ so that the data in $2.0 < U < 2.4$ are
for metallic region.  Thus, they are not directly compared with the
present results in the insulator regime.  The dynamical CPA+HA result at
$U=2.4$ agrees with the QMC result.  These results suggest that 
the HA gives the quantitative results of 
$\langle n_{\uparrow}n_{\downarrow} \rangle$ in both the weak 
($U < 1.0$) and strong ($U > 2.4$) Coulomb interaction regimes.
In the intermediate Coulomb interaction regime 
($1.0 \lesssim U \lesssim 2.3$), we need to take into account more 
dynamical effects to obtain   
$\langle n_{\uparrow}n_{\downarrow} \rangle$ quantitatively 
at low temperatures.
%---------------------------------------------------------------------
\begin{figure}
\includegraphics[scale=0.7]{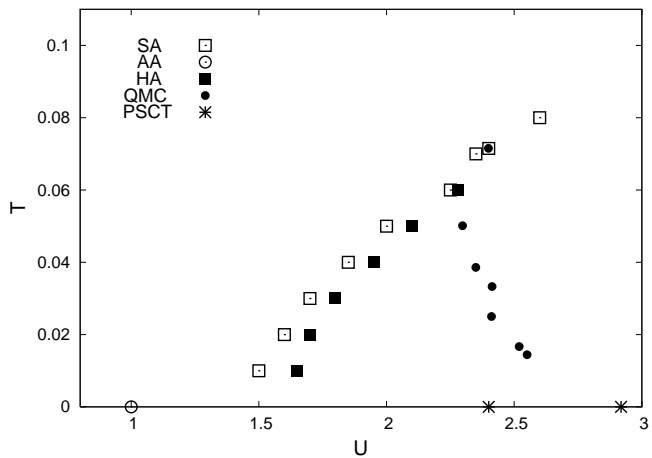}%
\caption{\label{figuc}
Phase boundaries of metal-insulator transition below $T_{\rm N}$
in the SA (open squares), the HA (closed squares), and 
the QMC method (closed circles)~\cite{schl99}.
The critical Coulomb interaction in the alloy-analogy approximation 
(AA)~\cite{hub63}
 at $T=0$ is indicated by the open circle, while the exact results of
 $U_{\rm c1}$ and $U_{\rm c2}$ at $T=0$ based on the projective
 self-consistent technique (PSCT)~\cite{moe95} are indicated by the stars. 
}
\end{figure}
%---------------------------------------------------------------------

We have plotted in Fig. 7 the MI transition points below $T_{\rm N}$
being obtained from the gap formation in the DOS.  
The critical Coulomb interactions $U_{\rm c}$ for gap formation in the 
full SA increase with increasing temperature, and merge into the QMC
results~\cite{schl99} around $T \approx 0.06$.  Since the full SA is 
exact in the high temperature limit, this indicates that the full SA 
quantitatively describes $U_{\rm c}$ at $T > 0.06$.
It should be noted that these data are smoothly connected to the
data points above $T_{\rm N}$ for a gap formation in Fig. 3.  
Below $T=0.06$, the  $U_{\rm c}$ in the SA deviate from the QMC 
results, and reduces to $U_{\rm c}=1.0$ in the Hubbard
alloy-analogy approximation~\cite{hub63} at $T=0$.

The dynamical results of $U_{\rm c}$ in Fig. 7 are obtained 
by adding the dynamical correction 
$\Delta U_{\rm c} = U_{\rm c}({\rm HA}) - U_{\rm c}({\rm SA})$ in the
decoupling approximation to $U_{\rm c}({\rm SA})$ in the full 
SA~\cite{kcmmt}.  
It is known that there are two critical Coulomb
interactions at $T=0$, $U_{\rm c1}$ for the gap formation and 
$U_{\rm c2}$ for the disappearance of the quasiparticle 
state~\cite{geor96}.   
But we did not find numerically two solutions near $U_{\rm c}$ 
at finite temperatures ($T > 0.01$).  
The critical Coulomb interactions in the HA are larger than those of the
full static approximation.  The difference between the two becomes smaller
with increasing temperature and vanishes at $T > 0.06$ as it should be.
The critical Coulomb interaction extrapolated to the zero
temperature is about 1.6 in the present HA, 
while the exact values based on the projective self-consistent
technique (PSCT)~\cite{moe95} are reported to be $U_{\rm c1}=2.40$
and  $U_{\rm c2}=2.84$.  
One has to take into account the dynamical terms
higher than $U^{4}$ more seriously in order to describe the MI
transition quantitatively.
%---------------------------------------------------------------------
\begin{figure}
\includegraphics[scale=0.75]{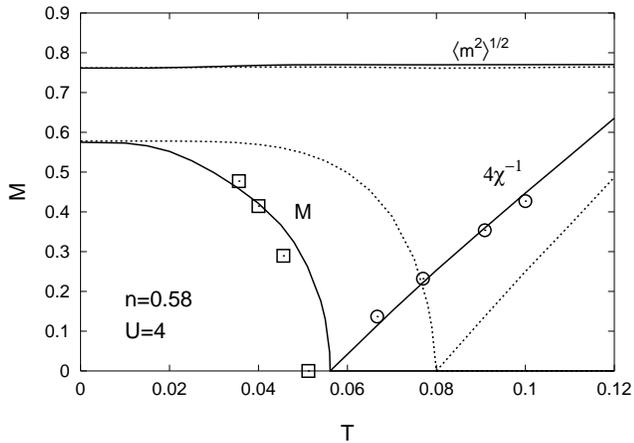}%
\caption{\label{figmtn0.58}
Magnetization ($M$), inverse susceptibility ($\chi^{-1}$) and 
amplitude of local moment ($\langle m^{2} \rangle^{1/2}$) 
as a function of temperature ($T$) on the fcc
 lattice in infinite dimensions.  Solid curves: dynamical CPA+HA, dotted
 curves : dynamical CPA+SA. Open squares (open circles) are the 
magnetizations (inverse susceptibilities) in the QMC~\cite{ulmke98}.
}
\end{figure}
%---------------------------------------------------------------------

\subsection{Ferromagnetism on the fcc lattice in infinite dimensions}
We have investigated the ferromagnetic properties of the fcc
lattice in infinite dimensions by means of the dynamical CPA+HA in order to
examine the quantitative aspects of the theory.
The fcc lattice in $d$ dimensions is defined as a lattice with 
$2d(d-1)$ nearest neighbors expressed by 
$\boldsymbol{R}= \pm \boldsymbol{e}_{i} \pm \boldsymbol{e}_{j}$ 
with two different cubic unit vectors $\boldsymbol{e}_{i}$ and 
$\boldsymbol{e}_{j}$ ($i, j = 1, 2, \cdots, d$).  In infinite dimensions
the noninteracting DOS is given by~\cite{mull91} 
\begin{eqnarray}
\rho(\epsilon) = 
\dfrac{{\rm e}^{\displaystyle -\frac{(1+\sqrt{2}\epsilon)}{2}}}
{\sqrt{\pi (1+\sqrt{2}\epsilon)}} \ .
\label{fccdos}
\end{eqnarray}
Here the energy unit is chosen so that the bottom of band edge is given
by $-1/\sqrt{2}$.  

Note that the DOS (\ref{fccdos}) monotonically increases with
decreasing energy $\epsilon$ and diverges at the band edge 
$\epsilon=-1/\sqrt{2}$.  Therefore the DOS is expected to be favorable 
for the ferromagnetism in the low density region.
Since the numerator in the DOS is reduced by a factor of 1/e at 
$\epsilon=1/\sqrt{2}$, 
we may regard $W=\sqrt{2}$ as a characteristic band width of this system.
The coherent Green function $F_{\sigma}(i\omega_{l})$ was calculated
from eq. (\ref{cohgp}) via numerical integration with
respect to energy $\epsilon$.

Figure 8 shows an example of calculated magnetization vs temperature
curves in various approximations for electron number $n=0.58$ and Coulomb
interaction $U=4.0$.  Extrapolated ground-state magnetizations are
almost saturated in both calculations.  
The magnetization vs. temperature curves considerably deviate downwards 
from the Brillouin curve. 
The SA overestimates the Curie temperature 
($T_{\rm C} = 0.0800$) as compared with the QMC~\cite{ulmke98} 
($T_{\rm C} = 0.0512$).  
The dynamical effects based on the HA reduces
$T_{\rm C}$ by 30\% and results in $T_{\rm C} = 0.0559$.  The result agrees
with the QMC~\cite{ulmke98} ($T_{\rm C} = 0.0512$) within 9\% error.
We have also calculated the paramagnetic susceptibilities above 
$T_{\rm C}$ by adding the infinitesimal magnetic field.  
Calculated susceptibilities follow the Curie-Weiss law; 
$\chi = m_{\rm eff}^{2}/(T - T_{\rm C})$.  Here $m_{\rm eff}$ is a
constant called effective Bohr magneton number.  Calculated
effective Bohr magneton numbers are 1.00 $\mu_{\rm B}$ in the SA, 1.13
$\mu_{\rm B}$ in the HA, and 1.18 $\mu_{\rm B}$ in the QMC,
respectively.  We find a good agreement between the dynamical CPA+HA
results and the QMC ones.

We have calculated $T_{\rm C}$ as a function of electron number $n$
by extrapolating the inverse susceptibilities to lower temperatures at
various $n$.
The results are presented in Figs. 9 and 10 for an intermediate Coulomb
interaction strength $U=2$ and a strong Coulomb interaction $U=4$, 
respectively.  
In the SA, we obtain a finite value of $T_{\rm C}$ for
infinitesimal electron number $n$.  The $T_{\rm C}$ increases linearly
with increasing $n$, and shows a maximum $T_{\rm C}=0.053\ (0.083)$ at
$n_{\rm max}=0.60\ (0.70)$ for $U=2\ (U=4)$.  When electron number $n$
approaches to one,  $T_{\rm C}$ vanishes again at $n=0.93\ (0.98)$ for
$U=2\ (U=4)$.  The dynamical effects reduce $T_{\rm C}$, for example, by 
35 \% (25 \%) at $n_{\rm max}=0.60\ (0.70)$ for $U=2\ (U=4)$.
As shown in Figs. 9 and 10, the dynamical results quantitatively agree
with the QMC results around the quarter filled electron number 
({\it i.e.}, $0.4 \lesssim n \lesssim 0.6$) for both $U=2$ and $U=4$.
However, the $T_{\rm C}$ vs $n$ curves shift to the higher density
region by about $\Delta n=0.1$; calculated $T_{\rm C}$ in the low
density regime ($0 < n \lesssim 0.3$)) are underestimated, and those in the
half-filled regime ($0.7 \lesssim n < 1$) are overestimated.

The behavior of $T_{\rm C}$ for small $n$ is sensitive to the
approximation.  In the Hartree-Fock approximation, $T_{\rm C}$ should 
be finite for infinitesimal number of electron according to the Stoner
condition $\rho(0) U > 1$ because the
noninteracting DOS diverge at $n=0$.  This holds true even in the static
approximation because the SA reduces to the Hartree-Fock approximation
in the weak Coulomb interaction limit.  The QMC results suggest that
$T_{\rm C}$ vanish at finite electron number in the low density
region. 
There is no exact analysis on the low-density behavior of $T_{\rm C}$
for this system as far as we know.
The present calculations indicate that $T_{\rm C}$ vanishes at $n=0.2$ 
for $U=2$, and at $n=0.3$ for $U=4$.  
%---------------------------------------------------------------------
\begin{figure}
\includegraphics[scale=0.75]{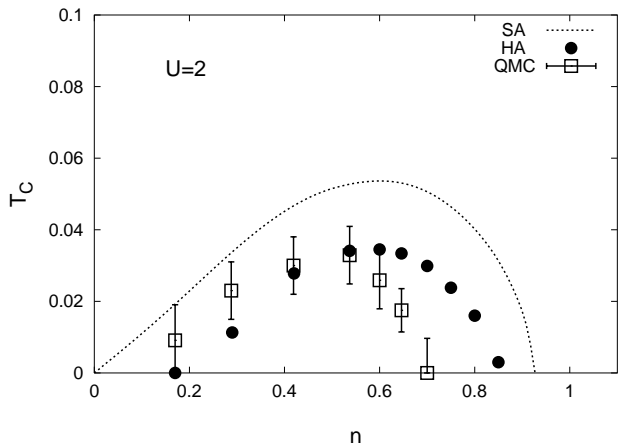}%
\caption{\label{figtc2}
Curie temperature ($T_{\rm C}$) vs. electron number ($n$) at $U=2$ in 
the SA (dotted curve), 
the HA (closed circles), and the QMC (open squares with error bars
)~\cite{ulmke98}.
}
\end{figure}
%---------------------------------------------------------------------
%---------------------------------------------------------------------
\begin{figure}
\includegraphics[scale=0.75]{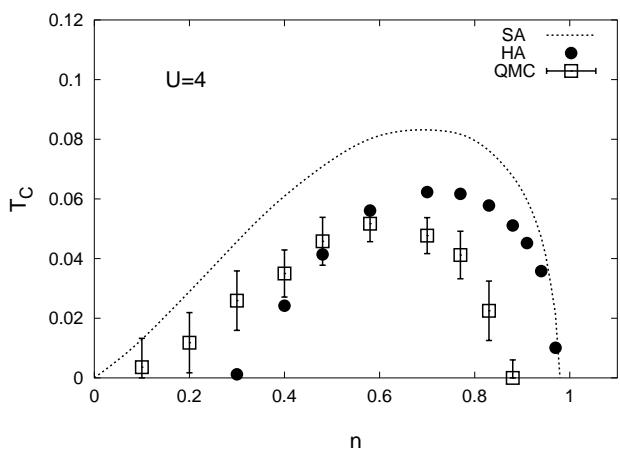}%
\caption{\label{figtc4}
Curie temperature vs. electron number at $U=4$ in the SA (dotted curve), 
the HA (closed circles), and the QMC (open squares with error bars
)~\cite{ulmke98}.
}
\end{figure}
%---------------------------------------------------------------------

Magnetic properties at the high temperatures above $T_{\rm C}$ are
expected to be described well by the dynamical CPA+HA.  We present in
Figs. 11 and 12 calculated effective Bohr magneton number 
($m_{\rm eff}$) and amplitude of local moment 
($\langle \mbox{\boldmath$m$}^{2} \rangle^{1/2}$) as a function of 
electron number $n$.  
Note that $\langle \mbox{\boldmath$m$}^{2} \rangle^{1/2} = \sqrt{3} 
\langle m^{2} \rangle^{1/2}$ where $\langle m^{2} \rangle^{1/2}$ is the
amplitude of local moment for $z$ component given by eq. (\ref{m23}).
These quantities were obtained at high temperatures ($T \approx 0.09$). 
In the case of $U=2$, the effective Bohr magneton number monotonically
increases with increasing electron number from 0 to 1, and shows the
maximum 1.64 $\mu_{\rm B}$ at half-filling.  
Although $m_{\rm eff}$ agrees with the amplitude 
$\langle \mbox{\boldmath$m$}^{2} \rangle^{1/2}$ in the atomic limit,  
it is in general smaller than the latter in the metallic state.
At half-filling, the difference is small; 
$m_{\rm eff} \approx \langle \mbox{\boldmath$m$}^{2} \rangle^{1/2}$ because of
considerable suppression of charge fluctuations.
By comparing the effective Bohr magneton numbers in the HA with those in
the SA, we find that the dynamical effects enhance the effective Bohr
magneton number, though the SA describes $m_{\rm eff}$ quantitatively at
half-filling. 

For larger value of $U=4$, we find similar behavior of $m_{\rm eff}$ as a
function of $n$, but the effective Bohr magneton number shows a maximum
at $n \approx 0.95$, and jumps from 1.3 $\mu_{\rm B}$ to 1.6 
$\mu_{\rm B}$ at $n=1$.  Sudden jump to the value 1.6 $\mu_{\rm B}$
being close to the atomic value $\sqrt{3}$ is associated with the formation
of an insulator at half-filling.  Calculated effective Bohr magneton
number at $n=0.58$ quantitatively agrees with the QMC result as shown in
Fig. 12.
%---------------------------------------------------------------------
\begin{figure}
\includegraphics[scale=0.75]{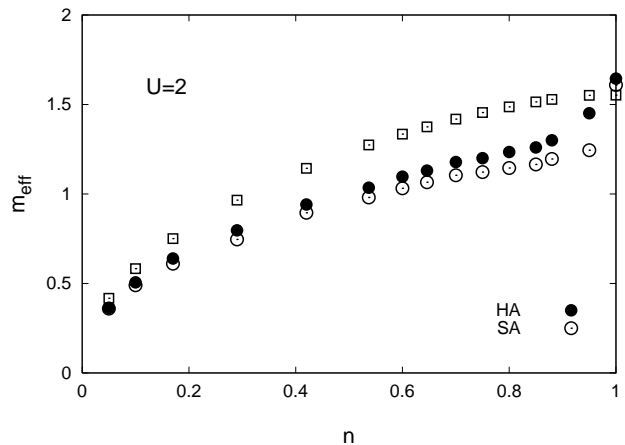}%
\caption{\label{figmeff2}
Calculated effective Bohr magneton numbers ($m_{\rm eff}$) 
as a function of the electron number $n$ in the HA (closed circles)
 and in the SA (open circles). Coulomb interaction energy is fixed to be
 $U=2$.  The amplitudes of local moments 
$\langle {\bf m}^{2} \rangle^{1/2}$ are also presented by open squares.
}
\end{figure}
%---------------------------------------------------------------------
%---------------------------------------------------------------------
\begin{figure}
\includegraphics[scale=0.75]{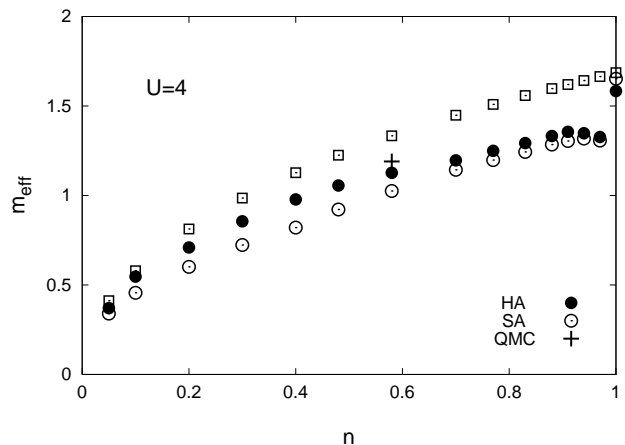}%
\caption{\label{figmeff4}
Same as in Fig. 11 but for $U=4$.
The QMC result at $n=0.58$ is shown by $+$.
}
\end{figure}
%---------------------------------------------------------------------

\section{Summary}
In the present paper we have investigated magnetic and electronic
properties of the Hubbard model in infinite dimensions on the basis of
the dynamical CPA combined with the harmonic approximation (HA) 
in order to clarify the accuracy of the theory at finite temperatures.

In the dynamical CPA, we transform the interacting electron system into
an independent electron system coupled with the time-dependent random
charge and exchange fictitious fields on the basis of the functional
integral method.  Introducing the coherent potential, we made a
single-site approximation in the free energy.  The coherent potential is
self-consistently determined by the condition that the average of 
dynamical impurity Green function 
embedded in the effective medium ({\it i.e.}, the
coherent potential) should be equal to the coherent Green function for
the uniform medium.  The former Green function is calculated by means 
of the HA.
The dynamical CPA+HA becomes exact in the high-temperature limit,
leads to the free energy being exact up to the second order in the weak
$U$ limit, and yields the exact result in the strong $U$ limit 
in infinite dimensions.

We have performed the numerical calculations with use of the dynamical
CPA+HA in which the dynamical corrections have been taken into account
up to the 4th order in $U$ exactly and the higher-order corrections 
up to the 16th order in $U$ have been taken into account with use of the 
second order asymptotic approximation.
In the present calculations for the half-filled band Hubbard 
model on the Bethe lattice in infinite dimensions we found that 
the N\'eel temperatures $T_{\rm N}$ 
are well described by the zeroth approximation to the dynamical CPA, 
{\it i.e.}, the static approximation (SA).  The dynamical
corrections based on the HA reduce further $T_{\rm N}$ in the weak
Coulomb interaction regime, leading to a quantitative agreement with the
QMC results.  In the intermediate Coulomb interaction regime, the 
dynamical corrections in the present approximation
tend to underestimate $T_{\rm N}$ by about 10 \%. 
In the strong Coulomb interaction regime we found that the dynamical 
corrections slightly increase $T_{\rm N}$.
These results suggest that the dynamical CPA+HA can describe $T_{\rm N}$
within 10\% error in infinite dimensions.

We have also verified using the same model at half filling that   
the SA quantitatively describes the MI crossover 
above $T_{\rm N}$.  It however underestimates the
critical Coulomb interaction $U_{\rm c}$ at low temperatures.  The
dynamical corrections based on the HA enhance $U_{\rm c}$.  But the
corrections in the present version are not enough to reproduce 
quantitatively 
the critical $U_{\rm c}$ obtained by the QMC; the present 
theory leads to $U_{\rm c}$ smaller than those of the QMC by 30\% 
at low temperatures ($T \sim 0.02$).

We have investigated the ferromagnetic properties on the fcc lattice 
in infinite dimensions.  
In this case, the SA overestimates the Curie temperatures $T_{\rm C}$ 
more than 30\% irrespective of electron number.  
The dynamical CPA+HA reduces $T_{\rm C}$ in the SA, 
and quantitatively describes 
the Curie temperatures as well as the susceptibilities in the 
quarter-filled regime ($0.4 \lesssim n \lesssim 0.6$). 
However the $T_{\rm C}$ vs $n$ curves tend to shift to the higher density 
region by about $\Delta n=0.1$ as compared with those in the QMC 
calculations, so that $T_{\rm C}$ in the low density region 
($0 < n \lesssim 0.3$) are somewhat underestimated and those in the 
half-filled regime ($0.7 \lesssim n < 1$) are overestimated.
We found that the effective Bohr magneton number $m_{\rm eff}$ is
smaller than the amplitude 
$\langle \mbox{\boldmath$m$}^{2} \rangle^{1/2}$
in itinerant electron system except at half-filling where electrons tend
to localize due to electron correlations for the intermediate Coulomb
interaction strength.  The dynamical effects enhance
$m_{\rm eff}$ for non half-filled bands, while the corrections are
negligible at half-filling.

We have to perform the full CPA calculations without decoupling
approximation (\ref{cpa3}) in order to obtain more solid conclusions. 
Nevertheless,
the present results of calculations indicate that the dynamical CPA
combined with the HA is suitable for the quantitative calculations of 
magnetic properties
at finite temperatures as well as the MI transition at high temperature
region.  It is not enough for the quantitative description of the 
MI transition at low temperatures and for the quantitative description
of the ferromagnetism in the low-density regime as well as in the
half-filled regime.  For more quantitative description 
in the intermediate Coulomb
interaction and low-temperature regime, one has to treat more 
seriously the higher-order dynamical corrections.  The improvement 
is left for future work.

\section*{Acknowledgment}
One of the authors (T.T.) is grateful to Mr. T. Nakamura for valuable
discussions on the densities of states in infinite dimensions.
This work was partly supported by the Ministry of Education, 
Science, Sports and Culture, Grant-in-Aid for Scientific Research (C), 
22540395, 2010.
Numerical calculations have been partly carried out with use of the
Hitachi SR11000 in the Supercomputer Center, Institute of Solid State
Physics, University of Tokyo.


\begin{thebibliography}{99}
\bibitem{kcmmt}
Note that $\Delta T_{\rm N} (\Delta U_{\rm c})$ is not caused by the
	decoupling approximation, but is caused by the dynamical
	corrections in eqs. (\ref{edyn3}) and (\ref{effimpg}).
\end{thebibliography}
\end{document}